\documentclass[
twocolumn, 
 amsmath,amssymb,
 aps,prl,
superscriptaddress
]{revtex4-2}

\usepackage{graphicx}
\usepackage{dcolumn}
\usepackage{bm}
\usepackage[colorlinks=true, allcolors=blue]{hyperref}
\usepackage{braket}
\usepackage{bbding}
\usepackage{natbib}
\usepackage{gensymb}
\usepackage{balance}
\usepackage{placeins}
\usepackage{mathtools}

\begin{document}
\preprint{APS/123-QED}

\title{Competitive Orders in Altermagnetic Chiral Magnons}

\author{Congzhe Yan}
\affiliation{%
 Division of Physics and Applied Physics, School of Physics and Mathematical Sciences, Nanyang Technological University, Singapore 637371, Singapore 
}%
\affiliation{%
Department of Physics, University of Science and Technology of China, Hefei, Anhui, 230026, China 
}%

\author{Zhijun Jiang}
\affiliation{%
 Ministry of Education Key Laboratory for Nonequilibrium Synthesis and Modulation of Condensed Matter, Shaanxi Province Key Laboratory of Advanced Functional Materials and Mesoscopic Physics, School of Physics, Xi’an Jiaotong University, Xi’an 710049, China 
}%

\author{Jinyang Ni}
\email{jyni@xjtu.edu.cn}
\affiliation{%
 Ministry of Education Key Laboratory for Nonequilibrium Synthesis and Modulation of Condensed Matter, Shaanxi Province Key Laboratory of Advanced Functional Materials and Mesoscopic Physics, School of Physics, Xi’an Jiaotong University, Xi’an 710049, China 
}%

\author{Guoqing Chang}
\email{guoqing.chang@ntu.edu.sg}
\affiliation{%
 Division of Physics and Applied Physics, School of Physics and Mathematical Sciences, Nanyang Technological University, Singapore 637371, Singapore 
}%




\begin{abstract}
The magnons in altermagnets exhibit chiral splitting even in the absence of spin-orbit coupling and external magnetic fields. Typically, this chiral splitting behavior can be well described by alternating isotropic spin exchanges (ISE) near the zero temperature. However, its finite-temperature dynamics, particularly when incorporating spin–orbit coupling, remains elusive. In this study, we reveal that, when including magnon–magnon interactions, long-range anisotropic spin exchange (ASE) can effectively induce chiral splitting of magnons at finite temperatures. Crucially, the ASE-induced chiral splitting competes with that arising from ISE, leading to a pronounced temperature-dependent modulation. Moreover, this competition is intimately coupled to quantum spin fluctuations, and can reverse the spin current driven by the band splitting as temperature increases. Our work uncovers the intrinsic competition governing collective spin excitations in altermagnets, providing new insights into their finite-temperature dynamical behavior.
\end{abstract}
\maketitle

\begin{figure}[h]
\centering
\includegraphics[scale=0.525]{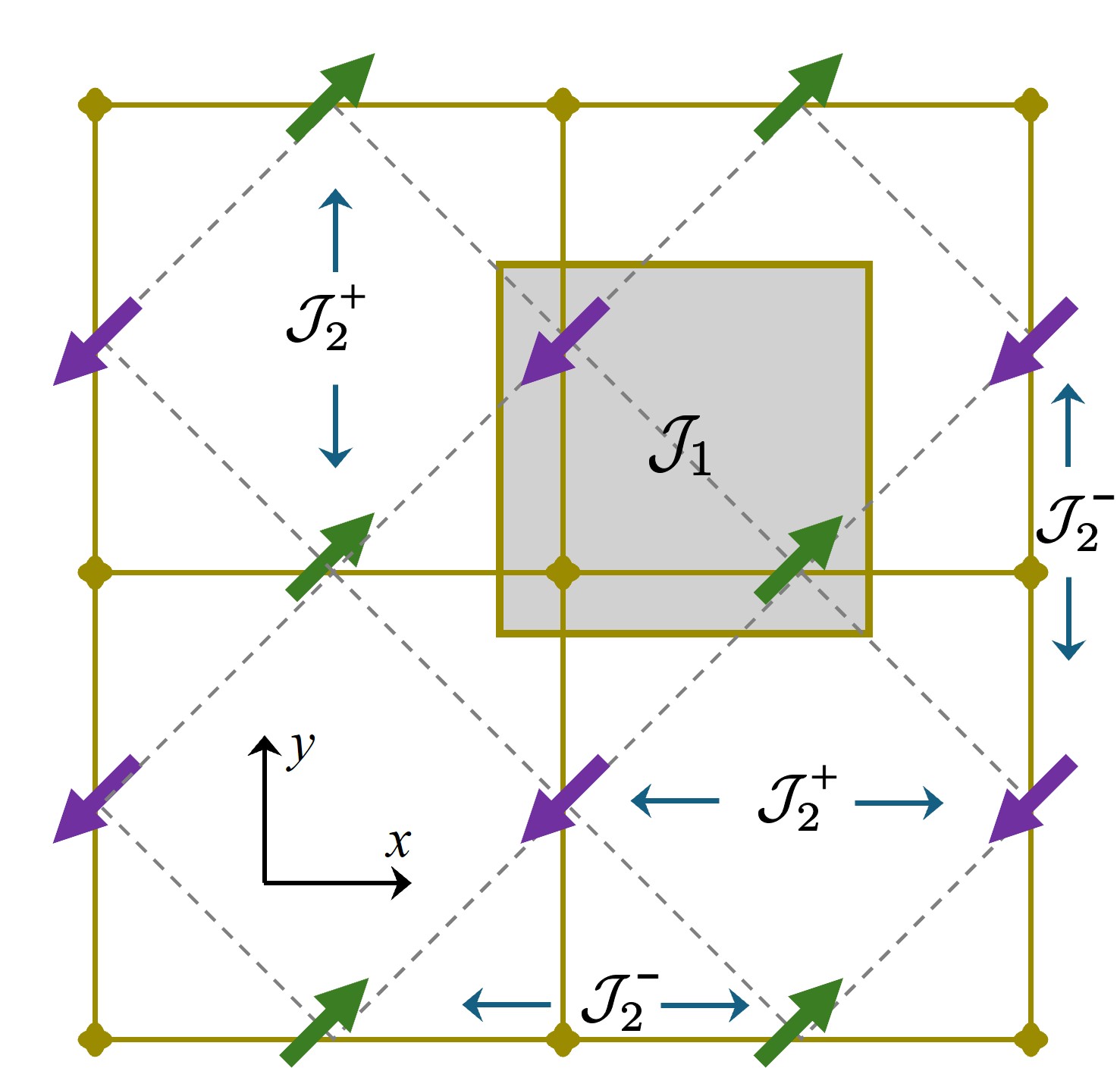}
\caption{The illustration of the $d$-wave altermagnet on the square lattice. 2NN spin exchanges exhibit the $d$-wave behavior (${\cal J}^{+}_{2}$\,$\neq$\,${\cal J}^{-}_{2}$) due to the breaking ${\cal PT\tau}$ symmetry between opposite-spin sublattices.}
\label{fig1}
\end{figure}

\textit{Introduction.} Conventional collinear antiferromagnets exhibit zero net magnetization and are preserved under translation\,(${\cal \tau}$) and inversion\,(${\cal P}$) symmetry, resulting in spin-degenerate bands\,\cite{jungwirth2016antiferromagnetic, baltz2018antiferromagnetic}. However, recent studies have shown that when opposite-spin sublattices are no longer linked by translation or inversion symmetry but instead by additional rotational or mirror symmetry operations\,\cite{vsmejkal2022emerging, vsmejkal2022beyond, hayami2019momentum}, such magnetic states can exhibit momentum-dependent spin-split bands even in the absence of spin-orbit coupling (SOC). This phenomenon has been experimentally confirmed in candidate materials, dubbed as altermagnets\,\cite{vsmejkal2020crystal, mazin2021prediction, bai2024altermagnetism, zhou2025manipulation, zhang2025electrical, duan2025antiferroelectric, song2025altermagnets, liu2022spin}. For example, in a $d$-wave altermagnet on a square lattice, the sublattices are related by a spin flip combined with a $\pi/2$ real-space rotation about a point on the dual lattice, as illustrated in Fig.\,\ref{fig1}. The ability to combine ferromagnetic-like spin splitting with zero net magnetization avoids adverse stray-field effects during ultrafast magnetic dynamics, offering significant potential for next-generation spintronics\,\cite{he2023nonrelativistic, chen2024enumeration, jiang2024enumeration, xiao2024spin, liu2024twisted,  das2024realizing, leeb2024spontaneous}.

Similar to electrons, magnons—the quanta of collective spin excitations in altermagnets—exhibit non-relativistic band splitting behavior\,\cite{vsmejkal2023chiral,cui2023efficient,hoyer2025spontaneous, liu2024chiral, alaei2025origin, sun2025observation}. Calculations based on linear spin wave theory (LSWT) demonstrate that the alternating isotropic spin exchange (ISE), governed by the combined spin and crystal symmetries in altermagnets, gives rise to the chiral magnon band splitting\,\cite{vsmejkal2023chiral, cui2023efficient, hoyer2025spontaneous, hoyer2025altermagnetic, sun2025observation}. Extensive efforts have been devoted to exploring possible altermagnetic magnon band structures, but recent experimental observations show that the detected band splitting is extremely weak, displaying a pronounced deviation from theoretical predictions\,\cite{morano2025absence}. This discrepancy likely originates from the fact that the conventional Heisenberg model within LSWT fails to account for SOC effects\,\cite{lado2017origin,sabani2025beyond, mook2021interaction}. More importantly, the influence of many-body interactions on spin excitations in altermagnets at finite temperature remains largely unexplored\,\cite{garcia2025magnon, eto2025spontaneous, cichutek2025quantum}. 

\begin{figure*}
\centering
\includegraphics[scale=0.45]{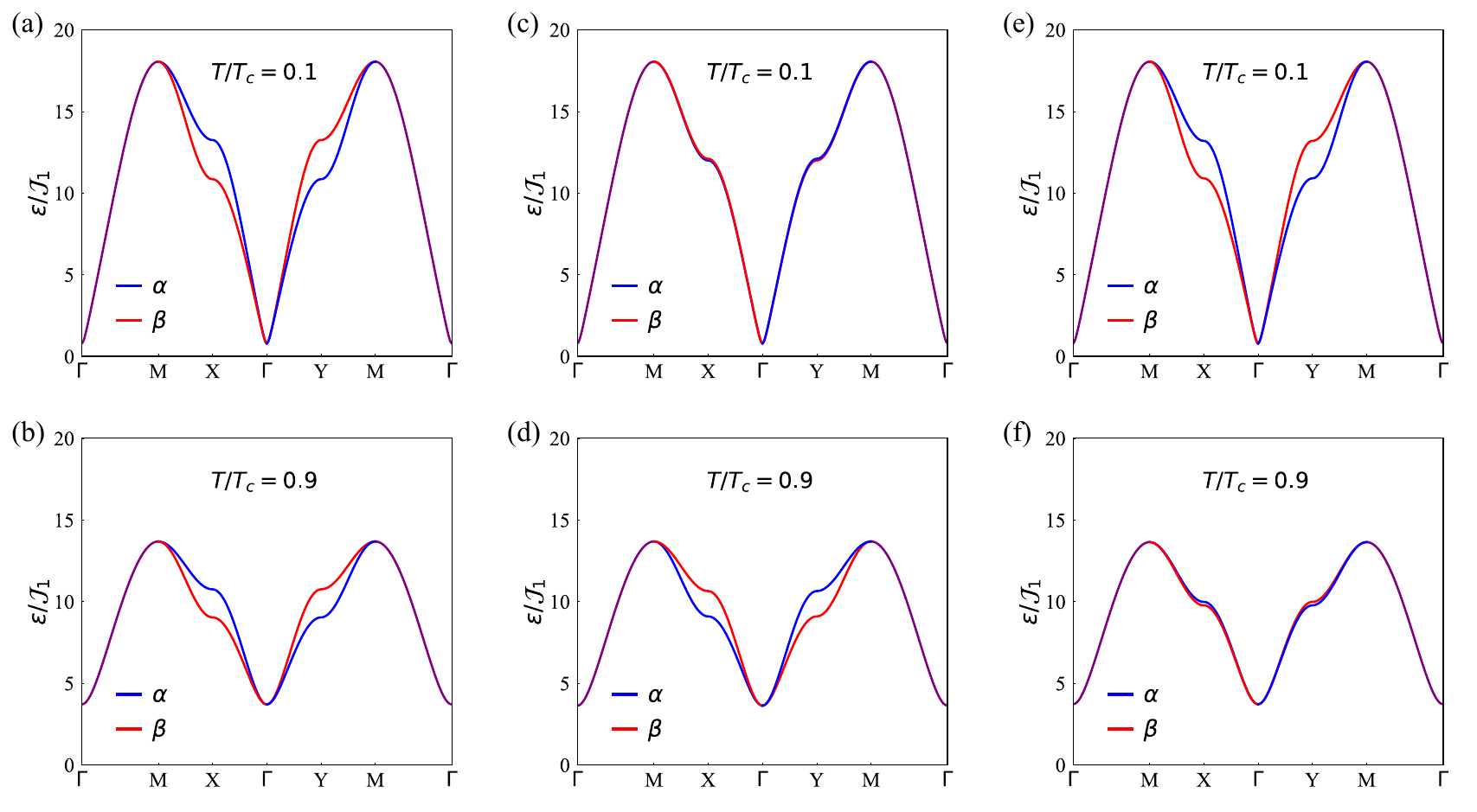}
\caption{Magnon bands of the monolayer $d$-wave altermagnet at different temperatures. (a)(b) With only ISE\,:\,$\Delta_{\cal J}/{\cal J}_{1}$\,$=$\,$0.2$. (c)(d) With only ASE\,:\,${\cal J}_{2z}/{\cal J}_{1}$\,$=$\,$0.6$. (e)(f) With both ISE\,:\,$\Delta_{\cal J}/{\cal J}_{1}$\,$=$\,$0.2$ and ASE\,:\,${\cal J}_{2z}/{\cal J}_{1}$\,$=$\,$0.6$.}
\label{fig2}
\end{figure*}

In this work, based on renormalized spin-wave theory (RSWT), we demonstrate that in $d$-wave altermagnets, long-range anisotropic spin exchange (ASE), though suppressed near the zero temperature, can effectively induce altermagnetic splitting of magnons at finite temperatures. Importantly, the altermagnetic splitting originating from ASE competes with that arising from alternating isotropic spin exchange (ISE), significantly modifying the overall altermagnetic magnon band structure as temperature increases. Moreover, this competing mechanism is strongly correlated with spin fluctuations, and can even reverse the sign of spin currents driven by the band splitting.

\textit{General spin model.} In this work, we consider a monolayer $d$-wave square altermagnet, as illustrated in Fig.\,\ref{fig1}. The minimal spin model is given by 
\begin{equation}\label{H0}
{\cal\hat H}={\cal\hat H}_{i}+{\cal\hat H}_{a}+{\cal\hat H}_{s}, 
\end{equation}
where ${\cal\hat H}_{i}$ includes nearest-neighbor (NN) and second-nearest-neighbor (2NN) ISE, ${\cal\hat H}_a$ corresponds to 2NN ASE, and ${\cal\hat H}_s$ describes the magnetic anisotropy which aligns the spins along the $z$-axis. In addition, for simplicity, we restrict our analysis to cases where the spin angular momentum $\langle{\cal S}_{z}\rangle$ is conserved.

The first part of the spin Hamiltonian, ${\cal\hat H}_{i}$, can be written as
\begin{equation}\label{H0i}
\begin{split}
{\cal\hat H}_{i} & = {{\cal J}_1}{\sum_{\langle{i,j}\rangle}}{{\bm S}_{\scriptstyle{i,\uparrow}}{\cdot}{\bm S}_{\scriptstyle{j,\downarrow}} }\\
& +{{\cal J}_2^+}{\sum_{\langle{i,j}\rangle^{\prime}_x}}{{\bm S}_{\scriptstyle{i,\uparrow}}{\cdot}{\bm S}_{\scriptstyle{j,\uparrow}}}+{{\cal J}_2^-}{\sum_{\langle{i,j}\rangle^{\prime}_y}}{{\bm S}_{\scriptstyle{i,\uparrow}}{\cdot}{\bm S}_{\scriptstyle{j,\uparrow}}}\\
& +{{\cal J}_2^-}{\sum_{\langle{i,j}\rangle^{\prime}_x}}{{\bm S}_{\scriptstyle{i,\downarrow}}{\cdot}{\bm S}_{\scriptstyle{j,\downarrow}}}+{{\cal J}_2^+}{\sum_{\langle{i,j}\rangle^{\prime}_y}}{{\bm S}_{\scriptstyle{i,\downarrow}}{\cdot}{\bm S}_{\scriptstyle{j,\downarrow}}},
\end{split}
\end{equation}
where the first term corresponds to NN ISE, and the other terms describe 2NN ISE. To maintain the collinear spin order, we impose that ${\cal J}_{1}$\,$>$\,$0$ and ${\cal J}_{2}$\,$=$\,$\frac{{\cal J}_{2}^{+}+{{\cal J}_{2}^{-}}}{2}$\,$\ll$\,${\cal J}_{1}$. In monolayer $d$-wave altermagnets, the lattice symmetry naturally gives rise to a nonzero $d$-wave $\Delta_{\cal J}$\,$=-$\,$\frac{{\cal J}_{2}^{+}-{\cal J}_{2}^{-}}{2}$.  ${\cal\hat H}_{s}$ represents the easy-axis single-ion anisotropy (SIA), described as ${\cal\hat H}_{s}$\,$=$\,${\cal K}\sum_{i}({\cal S}^{z}_{i})^{2}$ with ${\cal K}$\,$<$\,$0$. In the following discussion, we assume that $S$\,$=$\,$3/2$, ${{\cal J}_2}$\,$=$\,$-{{\cal J}_1}$, ${\cal K}$\,$=$\,$-0.01\,{{\cal J}_1}$.

Since $\langle{\cal S}_{z}\rangle$ is conserved and Dzyaloshinskii-Moriya Interaction (DMI)\,\cite{dzyaloshinsky1958thermodynamic, moriya1960anisotropic} is forbidden, 2NN ASE contains only diagonal exchange terms ${\cal J}_{2}^{xx}$, ${\cal J}_{2}^{yy}$ and ${\cal J}_{2}^{zz}$\,\cite{ASE_Note1}. In addition, the terms ${\cal J}_{2}^{xx}$, ${\cal J}_{2}^{yy}$ could be incorporated into the 2NN ISE term, allowing the ASE to be simplified as ${\cal J}_{2}^{zz}{S}_{i}^{z}{S}_{j}^{z}$\,\cite{ASE_Note1}. Therefore, in $d$-wave altermagnetic square lattice, ${\cal\hat H}_{a}$ is given as  
\begin{equation}\label{H0a}
\begin{split}
{\cal\hat H}_{a} & = {{\cal J}^{+}_{2z}}{\sum_{\langle{i,j}\rangle^{\prime}_x}}{S}_{\scriptstyle{i,\uparrow}}^{z}{\cdot}{ S}_{\scriptstyle{j,\uparrow}}^{z}+ {{\cal J}^{-}_{2z}}{\sum_{\langle{i,j}\rangle^{\prime}_y}}{S}_{\scriptstyle{i,\uparrow}}^{z}{\cdot}{ S}_{\scriptstyle{j,\uparrow}}^{z} \\
& + {{\cal J}^{-}_{2z}}{\sum_{\langle{i,j}\rangle^{\prime}_x}}{S}_{\scriptstyle{i,\downarrow}}^{z}{\cdot}{ S}_{\scriptstyle{j,\downarrow}}^{z}+{{\cal J}^{+}_{2z}}{\sum_{\langle{i,j}\rangle^{\prime}_y}}{S}_{\scriptstyle{i,\downarrow}}^{z}{\cdot}{ S}_{\scriptstyle{j,\downarrow}}^{z}, 
\end{split}
\end{equation}
Finally, to preserve the stability of the magnetic ground state, ${{\cal J}^{+}_{2z}}$\,$+$\,${{\cal J}^{-}_{2z}}$ must be equal to $0$. We denote ${\cal J}^{+}_{2z}$ by ${\cal J}_{2z}$ for simplicity.

\textit{Spin wave theory.} Next, we investigate the magnons of Eq.\,(\ref{H0}) based on the spin wave theory. Relying on a magnetically ordered classical ground state, we proceed with a Holstein-Primakoff (HP) transformation\,\cite{holstein1940field}
\newcommand{\opa}{\mathmakebox[0.6em]{a}}
\newcommand{\opb}{\mathmakebox[0.6em]{b}}
\newcommand{\vagger}{\vphantom{\dagger}}
\begin{equation}\label{HP}
\begin{split}
& \left\{\begin{aligned}
& S^{z}_{\scriptstyle{i,\uparrow}}=+S-{\opa}_{i}^{\dagger}{\opa}_{i}^{\vagger}\,({\hat n}_{i}^{\opa}) \\
& S^{+}_{\scriptstyle{i,\uparrow}}=\sqrt{2S-{\opa}_{i}^{\dagger}{\opa}_{i}^{\vagger}}{\opa}_{i}^{\vagger}\,{\approx}\,\sqrt{2S}{\opa}_{i}^{\vagger}-\frac{{\opa}_{i}^{\dagger}{\opa}_{i}^{\vagger}{\opa}_{i}^{\vagger}}{\sqrt{8S}} \\
& S^{-}_{\scriptstyle{i,\uparrow}}={\opa}_{i}^{\dagger}\sqrt{2S-{\opa}_{i}^{\dagger}{\opa}_{i}^{\vagger}}\,{\approx}\,\sqrt{2S}{\opa}_{i}^{\dagger}-\frac{{\opa}_{i}^{\dagger}{\opa}_{i}^{\dagger}{\opa}_{i}^{\vagger}}{\sqrt{8S}} \\
\end{aligned}\right. \\
& \left\{\begin{aligned}
& S^{z}_{\scriptstyle{j,\downarrow}}=-S+{\opb}_{j}^{\dagger}{\opb}_{j}\,({\hat n}_{j}^{\opb}) \\
& S^{+}_{\scriptstyle{j,\downarrow}}={\opb}_{j}^{\dagger}\sqrt{2S-{\opb}_{j}^{\dagger}{\opb}_{j}^{\vagger}}\,{\approx}\,\sqrt{2S}{\opb}_j^{\dagger}-\frac{{\opb}_{j}^{\dagger}{\opb}_{j}^{\dagger}{\opb}_{j}^{\vagger}}{\sqrt{8S}} \\
& S^{-}_{\scriptstyle{j,\downarrow}}=\sqrt{2S-{\opb}_{j}^{\dagger}{\opb}_{j}^{\vagger}}{\opb}_{j}^{\vagger}\,{\approx}\,\sqrt{2S}{\opb}_{j}^{\vagger}-\frac{{\opb}_{j}^{\dagger}{\opb}_{j}^{\vagger}{\opb}_{j}^{\vagger}}{\sqrt{8S}} \\
\end{aligned}\right. 
\end{split}
\end{equation}
At low temperatures ($T$\,$\ll$\,${\cal J}$), the perturbations about the classical ground state is very small ($\langle\hat{n}_{i,j}\rangle$\,$\ll$\,$S$)\,\cite{holstein1940field, zhitomirsky2013colloquium,ni2025magnon}. Therefore, for a low-temperature effective model, the higher-order interactions can be neglected. We retain only the ground-state energy and the quadratic terms in the bosonic operators, which together define as the first-order magnon Hamiltonian ${\cal \hat{H}}_{1}$. Notably, the quadratic bosonic terms arising from the ${\cal J}_{2z}$ spin exchanges is absent in the first-order magnon Hamiltonian, indicating that 2NN ASE makes no contribution to the LSWT approach.

We then employ Fourier transformation,
\newcommand{\bark}{{{\bar k}\vphantom{k}}}
\newcommand{\vark}{{{k}\vphantom{\bar k}}}
\begin{equation}\label{FT}
\begin{split}
& {\opa}_{\bm\vark}^{\vagger}=\frac{1}{\sqrt{N}}\sum_{i}{\opa}_{i}^{\vagger}e^{+i{\bm k}{\cdot}{\bm R}_i},\quad
{\opb}_{\bm\bark}^{\vagger}=\frac{1}{\sqrt{N}}\sum_{j}{\opb}_{j}^{\vagger}e^{-i{\bm k}{\cdot}{\bm R}_i}
\end{split}
\end{equation}
where $\bm\bark$ denotes magnons carrying momentum $-{\bm\vark}$. The first-order Hamiltonian can be expressed in the spinor basis $\psi^{\dagger}_{\bm k}$\,$=$\,$({\opa}_{\bm\vark}^{\dagger},{\opb}_{\bm\bark}^{\vagger})$ as ${\cal\hat H}_{1}$\,$=$\,$\sum_{\bm k}\psi_{\bm k}^{\dagger}{\cal\hat H}_{1\bm k}\psi_{\bm k}^{\vagger}$, with details shown in the supplemental material (SM)\,\cite{Supplemental_Materials}. Neglecting the zero-point energy, the ${\cal\hat H}_{1}$ in $\bm k$ space reads as
\begin{equation}\label{H1}
\frac{{\cal\hat H}_{1\bm k}}{S}=\lambda{I}+
\begin{pmatrix}
-{{\Delta}_{\cal J}}{f_d} & {{\cal J}_1}{f_\times} \\
{{\cal J}_1}{f_\times} & +{{\Delta}_{\cal J}}{f_d}
\end{pmatrix},
\end{equation}
where $\lambda$\,$=$\,$4{{\cal J}_1}$\,$-$\,$2{\cal K}$\,$+$\,${{\cal J}_2}{f_s}$ and the form factors $f_{\times}^{\bm k}$\,$=$\,$4$\,$\cos{\frac{k_x}{2}}$\,$\cos{\frac{k_y}{2}}$, $f_{x,y}^{\bm k}$\,$=$\,$2$\,$\cos{k_{x,y}}$\,$-$\,$2$, $f_{s,d}^{\bm k}$\,$=$\,$f_{x}^{\bm k}$\,$\pm$\,$f_{y}^{\bm k}$. Using Bogoliubov transformations\,\cite{bogoljubov1958new, valatin1958comments}, we can derive the analytical eigenvalues of Eq.\,(\ref{H1}), expressed as
\begin{equation}\label{E1}
\frac{{\cal E}_{\bm k}^{\alpha,\beta}}{S}=
\sqrt{{\left(4{{\cal J}_1}+{{\cal J}_2}{f_s}-2{\cal K}\right)}^2-{\left({{\cal J}_1}{f_\times}\right)}^2}\mp{\Delta_{\cal J}}f_d.
\end{equation}
It clearly shows that the emergence of $\mp{\Delta_{\cal J}}$ lifts the band degeneracy. 

\newcommand{\opalpha}{\alpha\vphantom{\beta}}
\newcommand{\opbeta}{\beta\vphantom{\alpha}}
As shown in Fig\,\ref{fig2}.(a), along the $\Gamma$-M(or M-$\Gamma$) path, the $\alpha$ and $\beta$ modes remain degenerate due to the $\lvert{k_x}\rvert$\,$=$\,$\lvert{k_y}\rvert$. When $\Delta_{\cal J}$\,$>$\,$0$, the $\alpha$ mode is higher in energy than $\beta$ mode along the M-X-$\Gamma$ path as $\lvert{k_x}\rvert$\,$>$\,$\lvert{k_y}\rvert$, while the opposite occurs along the $\Gamma$-Y-M path as $\lvert{k_x}\rvert$\,$<$\,$\lvert{k_y}\rvert$. This momentum-dependent spin splitting originates from the alternating spin exchanges\,\cite{vsmejkal2023chiral, cui2023efficient}, with the alternating splitting naturally coupled to the sign of $\Delta_{\cal J}$. The splitting of magnon bands between $\alpha$ and $\beta$ modes reaches its maximum at the X and Y points, defined as $\Delta_{\cal E}$\,$=$\,$\frac{{\cal E}_{X}^{\opalpha}-{\cal E}_{X}^{\opbeta}}{2}$\,$=$\,$\frac{{\cal E}_{Y}^{\opbeta}-{\cal E}_{Y}^{\opalpha}}{2}$. Here, $\Delta_{\cal E}$ is referred to as the band gap of altermagnetic magnons. To generalize, we introduce the ratio of this gap to the total bandwidth, $\phi$\,$=$\,$\frac{{\cal E}_{X}^{\opalpha}-{\cal E}_{X}^{\opbeta}}{{\cal E}^{+}-{\cal E}^{-}}$\,$=$\,$\frac{{\cal E}_{Y}^{\opbeta}-{\cal E}_{Y}^{\opalpha}}{{\cal E}^{+}-{\cal E}^{-}}$, which serves as the altermagnetic order parameter.

\begin{figure}
\centering
\includegraphics[scale=0.45]{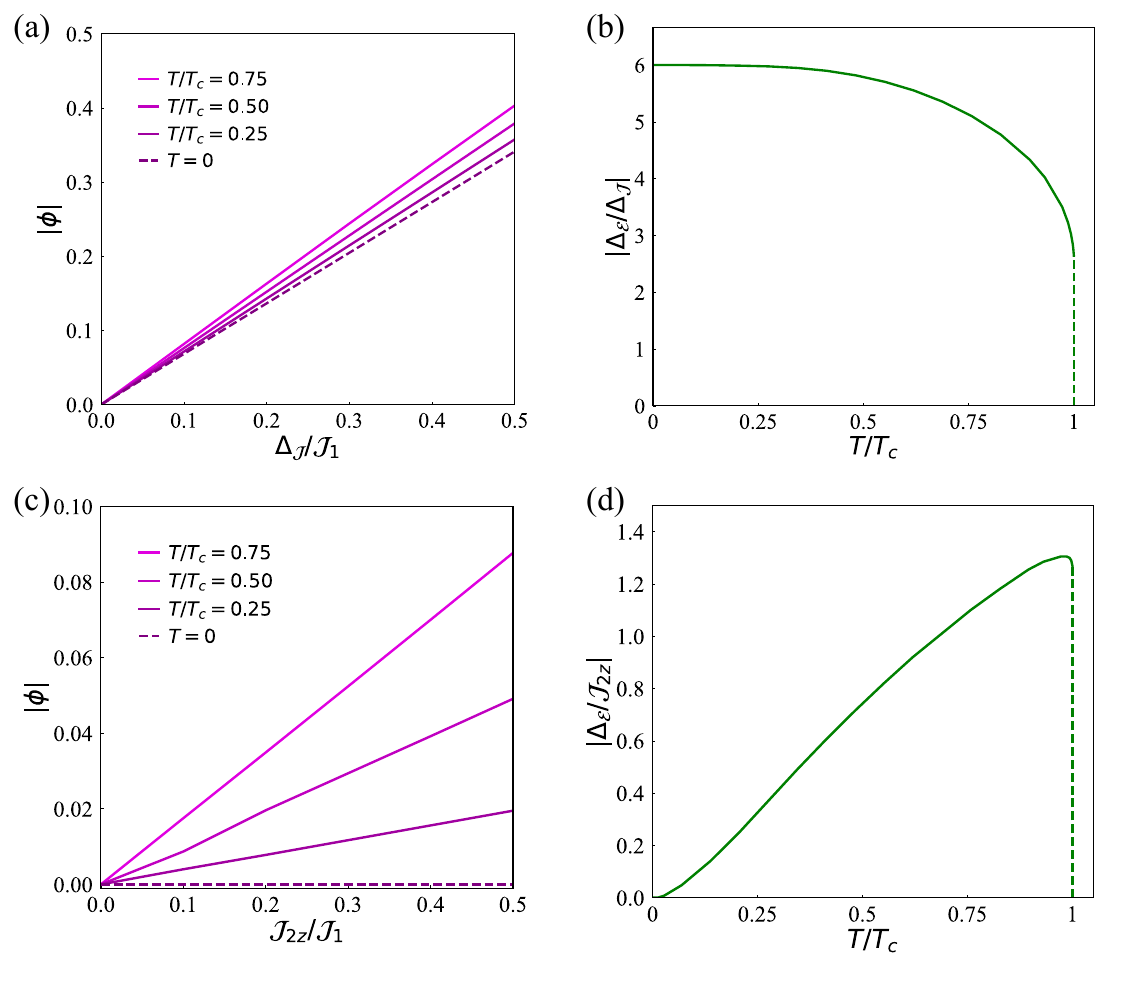}
\caption{Altermagnetic splitting of the $d$-wave altermagnet. (a)(c) Linear dependence of the order parameter $\phi$ on (a) ISE ($\Delta_{\cal J}/{\cal J}_{1}$) and (c) ASE (${\cal J}_{2z}/{\cal J}_{1}$) at a given temperature. (b)(d) Dependence of the linear coefficients for (b) ISE ($\Delta_{\cal E}/\Delta_{\cal J}$) and (d) ASE ($\Delta_{\cal E}/{\cal J}_{2z}$) on temperature $T$.}
\label{fig3}
\end{figure}

\textit{Many body effects.} At the high temperature regime ($T$\,$\sim$\,${\cal J}$), the magnon–magnon interactions cannot be neglected\,\cite{zhitomirsky2013colloquium, mook2021interaction, sourounis2024impact}. We extend the above spin Hamiltonian to include terms quartic in the bosonic operators, defined as the second order Hamiltonian ${\cal \hat{H}}_{2}$. Under the condition of momentum conservation, ${\bm k}_{1}$\,$+$\,${\bm k}_{2}$\,$=$\,${\bm k}_{3}$\,$+$\,${\bm k}_{4}$, the second order magnon Hamiltonian in $\bm k$ space can be expressed as
\begin{equation}\label{H2}
\begin{split}
 {\cal\hat H}_{2k} =& \frac{1}{N}\sum_{\{{\bm k}\}}
-{{\cal J}_1}f_{\times}^{{\bm q}^{\prime}}{\opa}_{{{\scriptstyle {\bm\vark}}}_1}^{\dagger}{\opb}_{{{\scriptstyle {\bm\bark}}}_3}^{\dagger}{\opb}_{{{\scriptstyle {\bm\bark}}}_2}^{\vagger}{\opa}_{{{\scriptstyle {\bm\vark}}}_4}^{\vagger} \\
& +\frac{1}{4}{{\cal J}_1}f_{\times}^{{\bm k}_1}{\opb}_{ {{\scriptstyle {\bm\bark}}}_1}^{\vagger}{\opa}_{{{\scriptstyle {\bm\vark}}}_2}^{\dagger}{\opa}_{{{\scriptstyle {\bm\vark}}}_3}^{\vagger}{\opa}_{{{\scriptstyle {\bm\vark}}}_4}^{\vagger}+{\mbox{H.c.}} \\
& +\frac{1}{4}{{\cal J}_1}f_{\times}^{{\bm k}_4}{\opb}_{ {{\scriptstyle {\bm\bark}}}_1}^{\dagger}{\opb}_{{{\scriptstyle {\bm\bark}}}_2}^{\dagger}{\opb}_{{{\scriptstyle {\bm\bark}}}_3}^{\vagger}{\opa}_{{{\scriptstyle {\bm\vark}}}_4}^{\dagger}+{\mbox{H.c.}} \\
& +\left({\cal M}_{\{{\bm k}\}}^{a}+{\cal K}\right){\opa}_{{{\scriptstyle {\bm\vark}}}_1}^{\dagger}{\opa}_{{{\scriptstyle {\bm\vark}}}_2}^{\dagger}{\opa}_{{{\scriptstyle {\bm\vark}}}_3}^{\vagger}{\opa}_{{{\scriptstyle {\bm\vark}}}_4}^{\vagger} \\
& +\left({\cal M}_{\{{\bm k}\}}^{b}+{\cal K}\right){\opb}_{{{\scriptstyle {\bm\bark}}}_1}^{\dagger}{\opb}_{{{\scriptstyle {\bm\bark}}}_2}^{\dagger}{\opb}_{{{\scriptstyle {\bm\bark}}}_3}^{\vagger}{\opb}_{{{\scriptstyle {\bm\bark}}}_4}^{\vagger},
\end{split}
\end{equation}
where ${{\bm q}^\prime}$\,$=$\,${{{\bm k}}_1}$\,$-$\,${{\bm k}}_4$, and the coefficient ${\cal M}_{\{\bm{k}\}}$ stems from 2NN spin exchanges, especially 2NN ASE, and is expressed as
\begin{equation}\label{M2}
\begin{split}
{\cal M}_{\{{{\bm k}}\}}^{a,b}=
& \pm\left({{\cal J}_{2z}}f_{\scriptstyle{d}}^{\scriptstyle{\bm q}}-\frac{\Delta_{\cal J}}{2}\big[2{f_{\scriptstyle{d}}^{\scriptstyle {\bm q}}}-f_{\scriptstyle{d}}^{\scriptstyle{{\bm k}}_1}-f_{\scriptstyle{d}}^{\scriptstyle{{\bm k}}_4}\big]\right) \\
& +\frac{{\cal J}_2}{2}\big[2{f_{\scriptstyle{s}}^{\scriptstyle{\bm q}}}-f_{\scriptstyle{s}}^{\scriptstyle{{\bm k}}_1}-f_{\scriptstyle s}^{\scriptstyle{{\bm k}}_4}\big],
\end{split} 
\end{equation} 
where ${\bm q}$\,$=$\,${{\bm k}_1}$\,$-$\,${ {\bm k}}_3$\,$=$\,${{{\bm k}}_4}$\,$-$\,${{\bm k}}_2$.

Next, we employ the Hartree-Fock (HF) approximation to deal with the magnon–magnon interactions. The HF approximation contracts the four-point interaction into the a sum of quadratic terms weighted by two-point correlation functions ${\cal\hat C}_{\bm k}$\,(for the formula of ${\cal\hat C}_{\bm k}$ see the SM\,\cite{Supplemental_Materials}). For instance, the first term of Eq.\,(\ref{H2}) could be approximated as\,\cite{li2018twodimensional, mkhitaryan2021self, wei2021renormalization}
\begin{equation}\label{HF}
\begin{split}
& \delta_{{{\bm k}_1+{\bm k}_2={\bm k}_3+{\bm k}_4}}\,{\opa}_{{\scriptstyle{\bm\vark}_1}}^{\dagger}{\opb}_{{\scriptstyle{\bm\bark}_3}}^{\dagger}{\opb}_{{\scriptstyle{\bm\bark}_2}}^{\vagger}{\opa}_{{\scriptstyle{\bm\vark}_4}}^{\vagger}
\approx \\
& \quad+\,\delta_{{{\bm k}_1={\bm k}_3},{{\bm k}_2={\bm k}_4}}
\left[\langle{\opa}_{{\scriptstyle{\bm\vark}_1}}^{\dagger}{\opb}_{{\scriptstyle {\bm\bark}_3}}^{\dagger}\rangle{\opb}_{{\scriptstyle {\bm\bark}_2}}^{\vagger}{\opa}_{{\scriptstyle {\bm\vark}_4}}^{\vagger} 
+\langle{\opb}_{{\scriptstyle{\bm\bark}_2}}^{\vagger}{\opa}_{{\scriptstyle{\bm\vark}_4}}^{\vagger}\rangle{\opa}_{{\scriptstyle{\bm\vark}_1}}^{\dagger}{\opb}_{{\scriptstyle {\bm\bark}_3}}^{\dagger}\right] \\
& \quad+\,\delta_{{{\bm k}_1={\bm k}_4},{{\bm k}_2={\bm k}_3}}
\left[\langle{\opa}_{{\scriptstyle{\bm\vark}_1}}^{\dagger}{\opa}_{{\scriptstyle {\bm\vark}_4}}^{\vagger}\rangle{\opb}_{{\scriptstyle{\bm\bark}_3}}^{\dagger}{\opb}_{{\scriptstyle{\bm\bark}_2}}^{\vagger}
+\langle{\opb}_{{\scriptstyle {\bm\bark}_3}}^{\dagger}{\opb}_{{\scriptstyle {\bm\bark}_2}}^{\vagger}\rangle{\opa}_{{\scriptstyle{\bm\vark}_1}}^{\dagger}{\opa}_{{\scriptstyle{\bm\vark}_4}}^{\vagger}\right].
\end{split}
\end{equation}
While the other terms could be approximated in a similar manner (for the details see the SM\,\cite{Supplemental_Materials}). Specifically, the 2NN ASE term after the HF approximation can be expressed as
\begin{equation}\label{H2*a}
{\cal\hat H}_{2,\mbox{\tiny HF}}^{a}=\frac{1}{N}\sum_{{\bm k},{\bm k}^{\prime}}
2{{\cal J}_{2z}}f_{d}^{{\bm k}^{\prime}-{\bm k}}
\left({\langle}{\hat n}_{{\bm \vark\prime}}^{\opa}{\rangle}{\hat n}_{{\bm\vark}}^{\opa}-{\langle}{\hat n}_{{\bm\bark\prime}}^{\opb}{\rangle}{\hat n}_{{\bm\bark}}^{\opb}\right).
\end{equation}
As a result, the overall effective Hamiltonian can be obtained by adding ${\cal\hat H}_{2,\mbox{\tiny HF}}$ to ${\cal\hat H}_{1}$, which only includes quadratic terms while effectively capturing the magnon-magnon interactions. 

\begin{figure}
\centering
\includegraphics[scale=0.465]{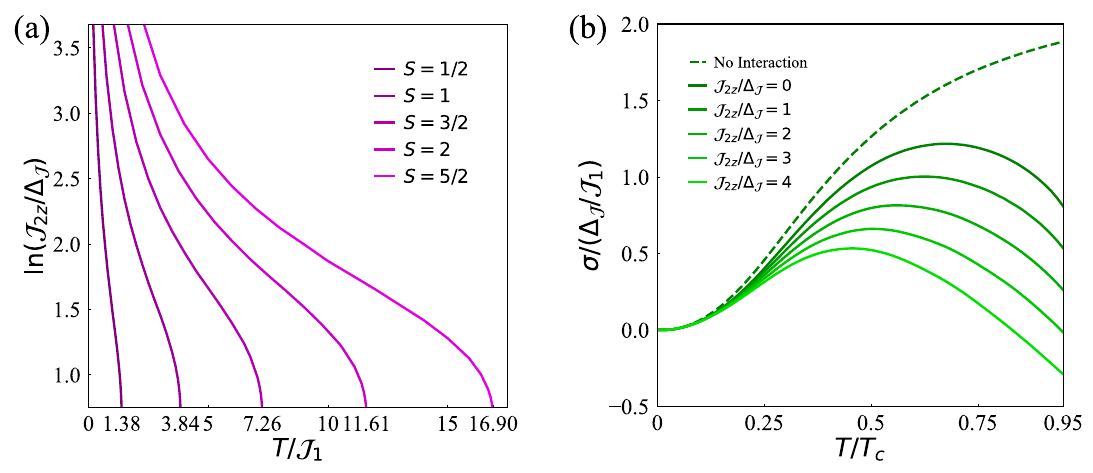}
\caption{(a) The dependence of the ISE/ASE ratio\,(${\cal J}_{2z}/\Delta_{\cal J}$) at which the altermagnetic gap vanishes completely on temperature $T$ and spin number $S$. (b) The dependence of thermal spin conductivity $\sigma$ on ratio ${\cal J}_{2z}/\Delta_{\cal J}$ and temperature $T$.}
\label{fig4}
\end{figure}

Based on the HF approximation, we calculate the renormalized magnon bands at finite temperatures. When only considering 2NN ISE, as shown in Fig.\,\ref{fig2}(a) and (b), increasing the temperature mainly modifies the shape of the magnon bands, while the altermagnetic splitting becomes slightly weaker, as 2NN ISE affects the energy of X/Y magnons directly and does not rely on other magnons. Interestingly, as shown in Figs.\,\ref{fig2}(c) and (d), the inclusion of 2NN ASE can also induce altermagnetic splitting at finite temperatures, which is significantly enhanced as the temperature increases. According to Eq.(\,\ref{H2*a}), ASE affects the bands by coupling the X/Y magnons with the $\Gamma$ magnons, and thus strongly depends on the magnon population at the $\Gamma$ point. For ${\cal J}_{2z}$\,$>$\,$0$, in the $\alpha$ band, ASE couples the X and $\Gamma$ magnons attractively and the Y and $\Gamma$ magnons repulsively, while in the $\beta$ band, the coupling is reversed. As a result, the sign of the altermagnetic splitting induced by 2NN ASE is opposite to that caused by 2NN ISE, and the altermagnetic splitting $\Delta_{\cal E}$ disappears or invert as the temperature rises [see Fig.\,\ref{fig2}(f) and Fig.\,S6\,\cite{Supplemental_Materials}]. In contrast, as illustrated in Fig.\,S5\,\cite{Supplemental_Materials}, when ${\cal J}_{2z}$ and ${\Delta}_{\cal J}$ possess opposite signs, ASE enhances the altermagnetic splitting with increasing temperature.

The temperature dependence of the order parameter $\phi$ and altermagnetic splitting $\Delta_{\cal E}$ are shown in Fig.\,\ref{fig3}. For a given temperature, $\phi$ exhibits a positive linear dependence on ISE. The linear coefficient $\Delta_{\cal E}/{\Delta}_{\cal J}$ varies with $T$ but changes slowly. In contrast, as shown in Figs.\,\ref{fig3}(c) and (d), order parameter $\phi$ also exhibits a linear dependence on ASE, with the linear coefficient $\Delta_{\cal E}/{\cal J}_{2z}$ varying almost linearly and rapidly with $T$. Therefore, compared with the ISE, the contribution of ASE becomes more significant at high temperature regime.

Given that magnon-magnon interactions is related to the spin quantum number $S$, it's essential to examine how the above competitive mechanism depends on $S$. As shown in Fig.\,\ref{fig4}(a), we determine the ratio between ${\cal J}_{2z}$ and $\Delta_{\cal J}$ required for the altermagnetic gap to vanish at different $S$. For altermagnets with a small $S$, strong spin fluctuations cause the altermagnetic gap to decrease rapidly with increasing temperature, so only a relatively small ${\cal J}_{2z}/{\Delta}_{\cal J}$ ratio is needed to fully close the gap. In contrast, in altermagnets with large $S$, the altermagnetic gap remains robust even at high temperature. This result provides important theoretical guidance for experimentally searching for robust altermagnetic magnons.

\begin{table}[b]
\caption{\label{tab1}DFT-calculated parameters and HF-calculated altermagnetic gaps in $d$-wave altermagnets $M\mbox{F}_{2}$ in which $M$\,$=$\,$\mbox{Ni}$,\,$\mbox{Co}$,\,$\mbox{Mn}$ (for more details see the SM\,\cite{Supplemental_Materials}). Here, {${\phi}_{t}$} refers to the altermagnetic order $\phi$ at $t$\,$=$\,$T/T_{c}$.}
\begin{ruledtabular}
\begin{tabular}{ccccccccc}
$M\mbox{F}_{2}$ & $S$ & ${\Delta}_{\cal J}$ & ${\cal J}_{2z}$ &{${\cal J}_{2z}/{\Delta}_{\cal J}$} & ${\phi}_{0.1}$ & ${\phi}_{0.3}$ & ${\phi}_{0.5}$ \tabularnewline
\hline
$\mbox{NiF}_{2}$ & $ 1 $ & $-0.219$ & $-0.011$ & $+0.050$ & $-0.264$ & $-0.321$ & $-0.296$ \tabularnewline
$\mbox{CoF}_{2}$ & $3/2$ & $-0.117$ & $+0.162$ & $-1.385$ & $-0.342$ & $-0.428$ & $-0.303$ \tabularnewline
$\mbox{MnF}_{2}$ & $5/2$ & $-0.043$ & $+0.001$ & $-0.023$ & $-0.073$ & $-0.083$ & $-0.066$ \tabularnewline
\end{tabular}
\end{ruledtabular}
\end{table} 

\textit{Spin currents.} The altermagnetic chiral splitting of magnons leads to a spin current driven by the temperature gradient\,\cite{cui2023efficient,wu2025magnon, onose2010observation, cheng2016spin, zyuzin2016magnon, ni2025magnon, sicheler2025optically}, known as the spin Seebeck effect. Given that spin conductivity is related to the splitting strength, the above competing mechanism between ASE and ISE has a significant impact on it. The thermal spin conductivity tensor $\sigma_{mn}$ is defined by $\langle{\bm{j}^z}\rangle_m$\,$=$\,$-\sigma_{mn}(\nabla{T})_n$, where $m,n$\,$\in$\,$\{x,y\}$, $\langle{\bm{j}^z}\rangle$ is the spin current density, $\nabla{T}$ refers to the applying temperature gradient. Based on the Kubo formula, the spin conductivity $\sigma_{mn}$ is given as 
\begin{equation}\label{SC}
\begin{split}
\sigma_{mn}=
& -\frac{\tau_{0}}{\hbar VT^{2}}{{\sum_{\bm k}}{\left({{\bm v}_{\bm k}^{\opalpha}}\right)}_m}{{\left({{\bm v}_{\bm k}^{\opalpha}}\right)}_n}{{\cal E}_{\bm k}^{\opalpha}}\frac{e^{{\cal E}_{\bm k}^{\opalpha}/T}}{\left(e^{{\cal E}_{{\bm k}}^{\opalpha}/T}-1\right)^2} \\
& +\frac{\tau_{0}}{\hbar VT^{2}}{{\sum_{\bm k}}{\left({{\bm v}_{\bm k}^{\opbeta}}\right)}_m}{{\left({{\bm v}_{\bm k}^{\opbeta}}\right)}_n}{{\cal E}_{\bm k}^\opbeta}\frac{e^{{\cal E}_{\bm k}^{\opbeta}/T}}{\left(e^{{\cal E}_{\bm k}^\opbeta/T}-1\right)^2},
\end{split}
\end{equation}
where ${\bm v}_{k}^{\alpha,\beta}$\,$=$\,${\partial\cal E}_{\bm k}^{\alpha,\beta}/{\hbar\partial{\bm k}}$, $\tau_0$ is the average magnon lifetime and $V$ is the volume. Note that $\tau_0$\,$=$\,$\hbar/{\cal J}_1$ and $k_B$\,$=$\,$1$ for simplicity. The thermal spin conductivity tensor exhibits $d$-wave behavior and could be simplified into a scalar $\sigma$\,$=$\,$+\sigma_{xx}$\,$=$\,$-\sigma_{yy}$ as $\sigma_{xy}$\,$=$\,$\sigma_{yx}$\,$=$\,$0$.

The calculated temperature dependence of thermal spin conductivity $\sigma$ is shown in Fig.\,\ref{fig4}(b). Since $\sigma$ depends linearly on altermagnetic splitting \,\cite{cui2023efficient, wu2025magnon}, it exhibits an almost linear dependence on $\Delta_{\cal J}$ at low temperatures, where magnon–magnon interactions are weak. As temperature increases, however, the magnon-magnon interactions is essential, the emergence of the ${\cal J}_{2z}$ narrow the altermagnetic splitting. Therefore, the giant ratio of ${\cal J}_{2z}/{\Delta}_{\cal J}$ can flip the sign of $\sigma$ at the high temperature regime. This intriguing result provides experimental guidance for detecting altermagnetic magnons in the 2D limit.

\textit{Discussion and summary}. The above results can be easily generalized from the $d$-wave altermagnet to other $f$-wave, $g$-wave and $i$-wave altermagnets, as well as from 2D materials to 3D materials. As shown in Tab.\,\ref{tab1}, we present the altermagnetic splitting of magnons in the $d$-wave altermagnets $M\mbox{F}_{2}$\,($M$\,$=$\,$\mbox{Ni}$,\,$\mbox{Co}$,\,$\mbox{Mn}$) at the finite temperatures\,\cite{morano2025absence, hoyer2025spontaneous}. This highlights that the ratio of ${\cal J}_{2z}/{\Delta}_{\cal J}$ play a decisive role in the altermagnetic splitting. For instance, the negative ratio of ${\cal J}_{2z}/{\Delta_{\cal J}}$ ensures that the altermagnetic gap in $\mbox{CoF}_{2}$ remain robust even for high temperatures. This suggest that ${\Delta_{\cal J}}$, ${\cal J}_{2z}$ and $S$ collectively determine the strength and thermal stability of the altermagnetic magnon splitting.

Importantly, this competitive mechanism between 2NN ASE and ISE is manifested in the spin currents induced by the band splitting. When ${\cal J}_{2z}$ and ${\Delta_{\cal J}}$ share the same sign, their altermagnetic splitting is opposite, and when ratio of ${\cal J}_{2z}/{\Delta_{\cal J}}$ is large, there is a reversal of the spin current at the high temperature regime. Conversely, when they have opposite signs, the competition turns into cooperation, thereby enhancing both the altermagnetic splitting and the thermal response of the spin current.

In summary, by developing RSWT, we provide an accurate description of altermagnetic magnons at finite temperature. Our results reveal that, beyond the alternating isotropic spin exchanges, the magnitude of altermagnetic splitting is governed by both by the long range ASE and spin fluctuations. These factors also influence the sign and magnitude of the spin current induced by the band splitting. Our findings provide a new picture for understanding altermagnetic magnons and provide theoretical guidance for the design of novel spintronics.

\textit{Acknowledgments.}
The authors thank Prof.\,Yuanjun Jin and Dr.\,Zhenlong Zhang for helpful discussions. This work was supported by the National Research Foundation, Singapore, under its Fellowship Award (NRFNRFF13-2021-0010); the Singapore Ministry of Education (MOE) Academic Research Fund Tier 3 grant (MOEMOET32023-0003); the Nanyang Assistant Professorship grant (NTU-SUG); USTC Fellowship (Grant No.\,U19582025), and  “Young Talent Support Plan” of Xi'an Jiaotong University.  

\textit{Data availability}. All data are available from the authors upon reasonable request.


\bibliography{main_ref}

\begin{thebibliography}{52}
\bibitem [{ASE()}]{ASE_Note1}%
  \BibitemOpen
  \href@noop {} {}\bibinfo {howpublished} {Since DMI interaction is forbidden due to the inversion symmetry, the 2NN ASE tensor is symmetric. The conservation of $\langle{{\cal S}_z}\rangle$ is directly leads to easy-axis anisotropy, which requires that the off-diagonal elements vanish, ${\cal J}_2^{xz}$\,$=$\,${\cal J}_2^{yz}$\,$=$\,${\cal J}_2^{xy}$\,$=$\,$0$, and in-plane diagonal elements be degenerate, ${\cal J}_2^{xx}$\,$=$\,${\cal J}_2^{yy}$. Consequently, the 2NN ASE term can be further simplified as ${\cal J}_2^{s}({\bm S}_{i}$\,$\cdot$\,${\bm S}_{j}$)\,$+$\,$({\cal J}_2^{zz}-{\cal J}_2^{s}){S}_{i}^{z}{S}_{j}^{z}$, where ${\cal J}_2^{s}$\,$=$\,${\cal J}_2^{xx}$\,$=$\,${\cal J}_2^{yy}$.}\BibitemShut {Stop}%
\end{thebibliography}

\end{document}